\documentclass[aip,apl,twocolumn,amsmath, amssymb]{revtex4}
\usepackage[T1]{fontenc}
\usepackage{amsmath,graphicx,amssymb,epsfig,babel,dsfont,amscd,ulem,tikz}
\usepackage{mathrsfs}
\usepackage{color}
\usepackage{placeins}

\renewcommand{\Im}{{\rm Im}}

\newcommand{\ri}{{\rm i}}

\newcommand{\kB}{k_{\rm B}}

\newcommand{\alphamat}{\underline{\underline{\alpha}}}
\newcommand{\chimat}{\underline{\underline{\chi}}}
\newcommand{\blockt}{\boldsymbol{T}^{-1}}

\begin{document}

\title{Dynamical thermal near-field routing with the non-reciprocal Weyl semi-metal Co$_3$Sn$_2$S$_2$}

\author{A. Naeimi}
\affiliation{Institut f\"{u}r Physik, Carl von Ossietzky Universit\"{a}t, 26111, Oldenburg, Germany}

\author{S.-A. Biehs}
\email{s.age.biehs@uol.de}
\affiliation{Institut f\"{u}r Physik, Carl von Ossietzky Universit\"{a}t, 26111, Oldenburg, Germany}

\date{\today}

\begin{abstract}
We demonstrate theoretically the non-reciprocal heating dynamics of two nanoparticles in the vicinity of a substrate all made of the ferromagnetic Weyl semi-metal Co$_3$Sn$_2$S$_2$. We show that the thermal routing effect is due to a spin-spin coupling mechanism between the nanoparticle resonances and the non-reciprocal surface modes of the substrate. Our numerical results indicate that the non-reciprocal heating effect is on the order of $22.5\%$ of the applied temperature differences. This strong rounting effect paves the way for first experimental realizations employing Weyl semi-metals and applications in nanoscale thermal management. 
\end{abstract}

\maketitle

%
%

Many-body systems involving magneto-optical materials offer the possibility to control the exchange of heat by thermal radiation at the nanoscale actively and passively. Typically by applying an external magnetic field, the time reversal symmetry is broken in these systems
which results in a breakdown of Lorentz reciprocity~\cite{Caloz}. As a consequence, thermal radiation can become non-reciprocal in the near and far-field regime which is manifested in different effects like a persistent heat current~\cite{Zhu2016}, a Hall effect for heat radiation~\cite{PBAHall2016}, spin and angular momentum~\cite{Silveirinha,Ottcircular2018,Zubin2019}, spin-directional thermal emission~\cite{DongEtAl2021}, inverse spin thermal Hall effect~\cite{PBA2025}, a giant magneto-resistance~\cite{Latella2017,Cuevas}, a Berry phase of thermal radiation~\cite{SABPBABerry2022}, strong heat flux rectification~\cite{Ottdiode2019,OttSpin2020} via non-reciprocal surface waves, and the Corbino effect~\cite{Latella2025} (see also reviews~\cite{OttReview2019,SABRMP2021}).

The application of strong magnetic fields at the nanoscale is, however, challenging and other methods
have been envisaged to realize non-reciprocal heat radiation by modulation techniques~\cite{FanMod2020,Floquet,SABGSA2023,RectMod}, for instance.
Another, even more practical possibility is to employ intrinsically non-reciprocal materials like 
non-reciprocal Weyl semi-metals (WSM)~\cite{ChenEtAl2016,Kotov2018,Guo_elight}. It could be shown that for such
materials one can have an anomalous photonic thermal Hall effect~\cite{OttAnomalous2020}, non-reciprocal surface wave induced heat flux rectification~\cite{HuEtAl2023,Naeimi2025}, modulated heat transfer in multilayer systems~\cite{TangEtAl2021,XuEtal2020,YuEtAlMultilayer2022,YuEtAl3B2022,Sheng2023}, negative differential thermal conductance~\cite{SunEtAl}, non-reciprocal emission and absorption~\cite{ZhaoEtAl2020,ZhangZhu2023}, broadband circularly polarized thermal radiation~\cite{Zubin}, non-reciprocal far-field heat radiation~\cite{WuEtAl2022}, and coupling to graphene nanoribbons~\cite{YuEtAl2023} (see also the reviews~\cite{GuoEtAl2023,DidariEtAl2024}). In particular, the possibility to enable thermal routing~\cite{GuoEtAl2020,GeEtAl2024,GeEtAl2025} by using the spin-spin coupling mechanism~\cite{OttSpin2020} offers many possibilites for an active and passive control of local radiative heat fluxes at the nanoscale. 

%
%

\begin{figure}
	\centering
	\includegraphics[width = 0.4\textwidth]{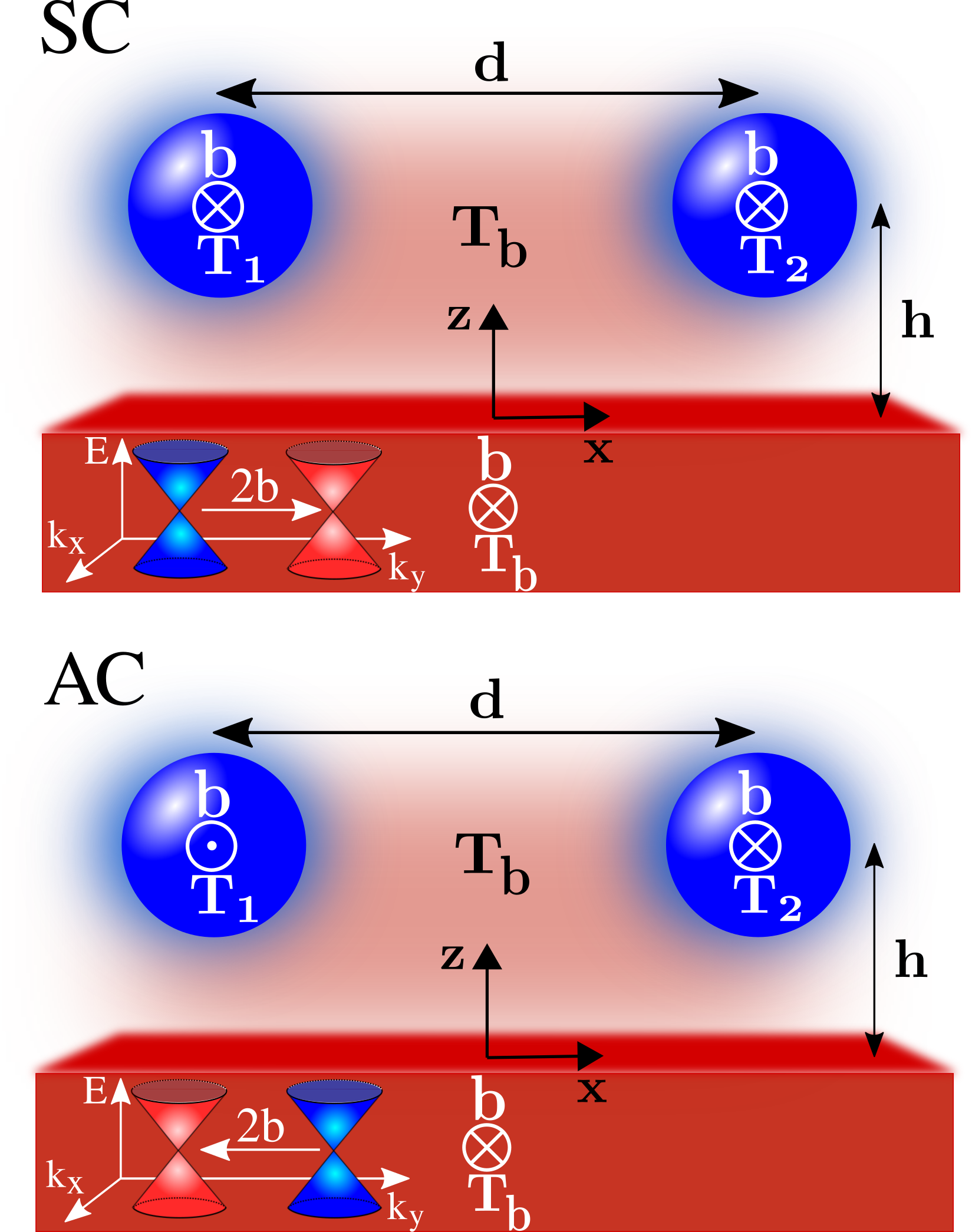}
	\caption{Sketch of the WSM router in symmetric configuration (SC) and asymmetric configuration (AC) of two WSM nanoparticles (NPs) above a substrate. The direction of the Weyl node seperation vector $2 \hat{b}$ is indicated in the NPs and in the substrate showing in $\pm y$ direction. The radii of the NPs are $R = 20\,{\rm nm}$ and $d = h = 5 R = 100\,{\rm nm}$. The initial temperatures are $T_b = 100\,{\rm K}$ and $T_1 = T_2 = 80\,{\rm K}$ and the heat flow from the substrate to the NPs is monitored.}
	\label{Fig:Sketch}
\end{figure}

In this letter, we investigate the dynamics of thermal routing between two WSM nanoparticles (NPs) and a WSM substrate as sketched in Fig.~\ref{Fig:Sketch}. We assume that all objects are made of Co$_3$Sn$_2$S$_2$ which is a ferromagnetic WSM with a magnetic order for temperatures below $T_c = 175\,{\rm K}$~\cite{heatcapacity} where it is non-reciprocal. Furthermore, we assume that the dipole approximation can be made. Then the polarizability tensor $\uuline{\alpha}$ of the WSM NPs is non-reciprocal, $\uuline{\alpha} \neq \uuline{\alpha}^t$, and one can expect that the three dipolar localized NP resonances with magnetic quantum numbers $m = 0,\pm1$ are non-degenerate due to a Zeeman splitting~\cite{Ottcircular2018,Raul1}. Furthermore, the permittivity tensor $\uuline{\epsilon}$ of the WSM substrate is non-reciprocal, i.e.\ $\uuline{\epsilon} \neq \uuline{\epsilon}^t$, and therefore the surface modes also undergo a Zeeman shift which leads to non-reciprocal properties as also known for magneto-optical materials~\cite{ChenEtAl2016,Kotov2018,Hofmann}. For the explicit expressions of the polarizability and permittivity tensor as well as the material parameters of Co$_3$Sn$_2$S$_2$ we refer the reader to Ref.~\cite{Naeimi2025}. From these explicit expressions it becomes apparent that for WSM the vector $2\mathbf{b}$ connecting the Weyl points plays formally the same role as the external magnetic field $\mathbf{B}$ for magneto-optical materials. Its direction determines the symmetry axis of the non-reciprocal polarizability and permeability and its magnitude determines the strength of the non-reciprocity. This allows us to define the symmetric configuration (SC) and asymmetric configuration (AC) by chosing different orientation of the Weyl node vector $2\mathbf{b}$. In the SC it was shown that the radiative heat flux between the NPs can have huge values of rectification ratios~\cite{HuEtAl2023,Naeimi2025}. Similarly for three NPs in an AC a routing effect has been highlighted~\cite{GuoEtAl2020} by studying the exchanged power between the three NPs. However, it is unclear how large the heating effect will be in such configurations, because the temperature change has not been investigated. Furthermore, when considering the heat flow between the NPs in the AC or SC as in Fig.~\ref{Fig:Sketch} by heating one of the NP and monitoring the heat flow to the second NP, there will be not only a heat flow between the NPs but also a substantial heat flow into the substrate. To avoid such a local heating of the substrate which is typically not taken into account in the theoretical approach, we consider the heat flux from the WSM substrate which is at the fixed temperature $T_b = 100\,{\rm K}$ to the WSM NPs which are initially held at $T_1 = T_2 = 80\,{\rm K}$. Then we study the exchanged power and the heating dynamics of the NPs towards the equilibrium state, i.e.\ the temperature change in the NPs, in order to demonstrate the non-reciprocal dynamical routing effect and its real impact on the temperature change. 

%
%

\begin{figure}
	\centering
	\includegraphics[width = 0.45\textwidth]{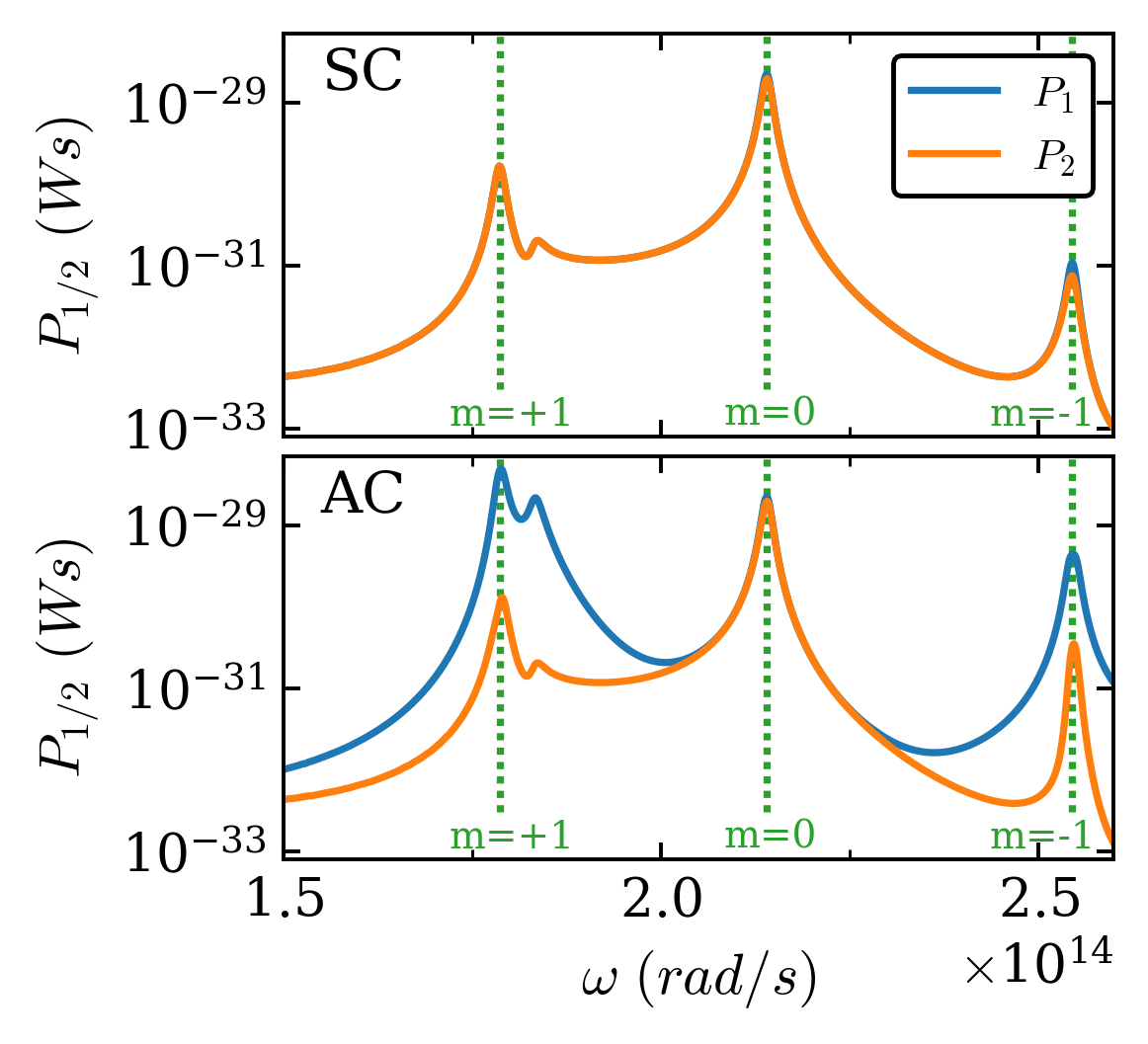}
	\caption{Spectral power $P_1(\omega)$ and $P_2(\omega)$ received by NP $1$ and $2$ for $T_1 = T_2 = 80\,{\rm K}$ and $T_b = 100\,{\rm K}$ for the SC and AC. The horizontal dashed lines mark the dipolar resonance frequencies $\omega_{m = 0, \pm 1}$ of the two NPs.}
	\label{Fig:P1P2}
\end{figure}

Now, to study the dynamical routing effect in the SC and AC we first determine the power received by the NPs. The power received by NP $1$  due to heat exchange with the substrate and with NP $2$ can be written in the form~\cite{OttSpin2020}
\begin{equation}
	\langle P_{1}\rangle = 3\int_{0}^{\infty}\frac{{\rm d}\omega}{2\pi}\hbar\omega \bigl[ (n_1 - n_b) \mathcal{T}_1^a + (n_2 - n_b) \mathcal{T}_1^b \bigr]
\label{p1}
\end{equation}
where $n_i = 1/(\exp(\hbar \omega/ \kB T_i) - 1)$ is the bosonic mean occupation number at the temperature $T_i$ with $i =1,2,b$. The transmission coefficients  $\mathcal{T}_1^a$ and $\mathcal{T}_1^b$ are relatively complicated and explicitely defined in Ref.~\cite{OttSpin2020}. In Fig.~\ref{Fig:P1P2} we show the spectral powers $P_1(\omega)$ and $P_2(\omega)$ in the SC and AC configuration. It can be nicely seen that there are three peaks corresponding to the three dipolar NP resonances at $\omega_{m = 0} = 2.1\times10^{14}\,{\rm rad/s}$, $\omega_{m = +1} = 1.76\times10^{14}\,{\rm rad/s}$, and $\omega_{m = -1} = 2.5\times10^{14}\,{\rm rad/s}$ which also exist in the absence of the substrate. Furthermore, it is obvious that the difference between $P_1(\omega)$ and $P_2(\omega)$ is negligible in the SC but very large in the AC. The main difference in the spectra for the AC is due to a strong heat flux difference at the frequencies of the two NP resonances at $\omega_{m = \pm 1}$. This difference will lead to a much stronger heat transfer to NP $1$ than to NP $2$ in the AC. To understand this effect, we first note that one can associate a thermal spin~\cite{OttSpin2020} to the NPs at both frequencies. This spin shows in the direction of the Weyl node separation vector $2\mathbf{b}$ for $\omega_{m = +1}$, i.e.\ in $y$ direction, and in the opposite direction for $\omega_{m = -1}$. Below we will see that the way this spin couples to the spin of the surface waves in the substrate allows us to understand the large heat flux difference for the AC.

%
%

\begin{figure}
	\centering
	\includegraphics[width = 0.45\textwidth]{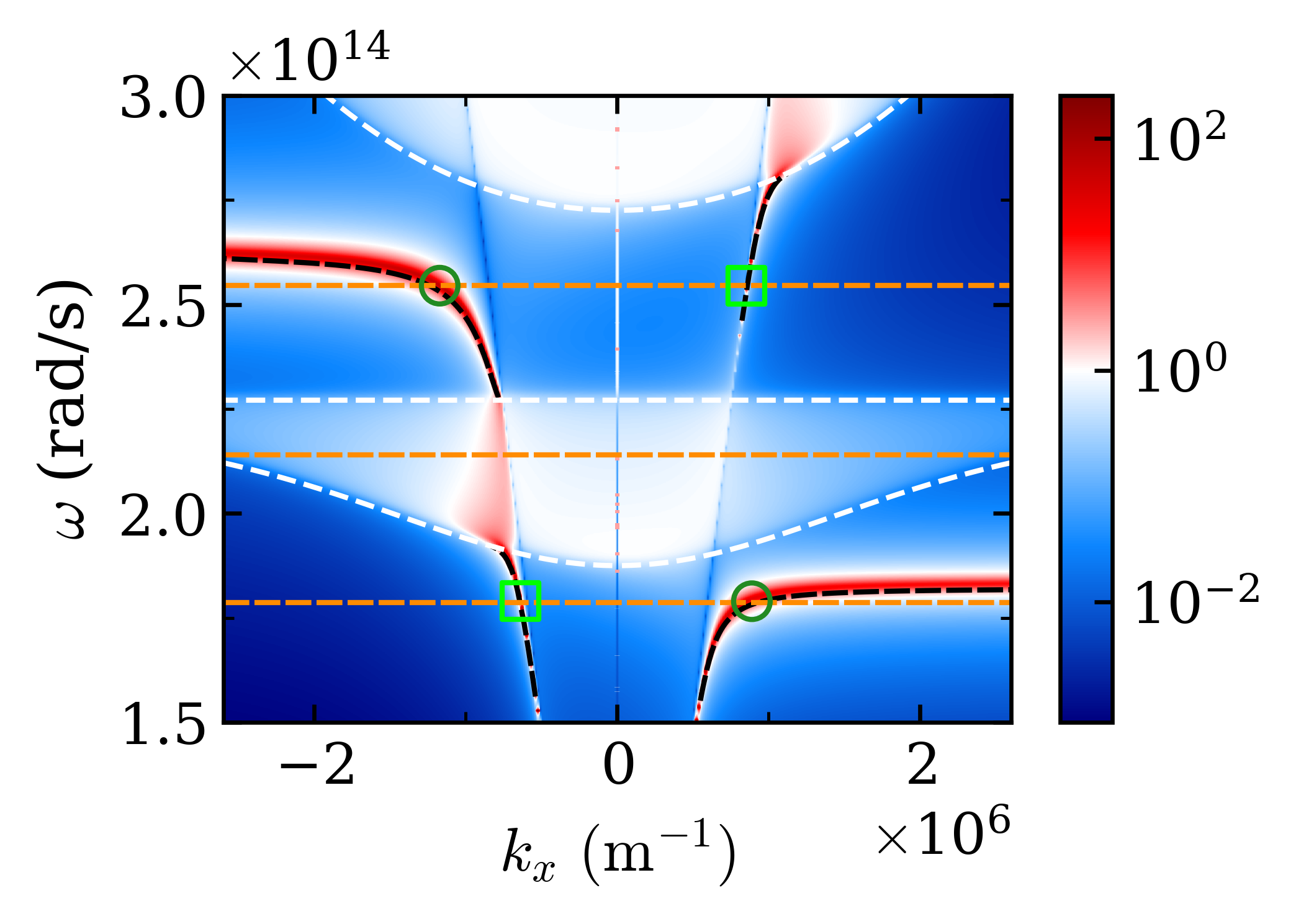}\\
	\caption{Plot of $1 -|r_{\rm pp}|^2$ for propagating waves ($|k_x| < \omega/c$) and $\Im(r_{pp})$ for evanescent waves ($|k_x| > \omega/c$) with $k_y = 0$ for the Co$_3$Sn$_2$S$_2$ substrate at temperature $T_b = 100\,{\rm K}$. The black dashed lines are the dispersion relations of the non-reciprocal surface modes obtained with the same method as in Ref.~\cite{Naeimi2025}. The horizontal lines indicate the three dipolar resonances of the  NPs at $T_1 = T_2 = 80\,{\rm K}$. The circles/boxes indicate the intersections of the localized NP resonances and the short range (circles) and the long-range (boxes) surface mode resonances which are spin-spin coupled to the corresponding NP resonances.}
	\label{Fig:ReflectionC}
\end{figure}

As detailed in Ref.~\cite{Kotov2018} and in Ref.~\cite{Naeimi2025} for WSM, the non-reciprocal substrate supports non-reciprocal surface waves. These surface waves are determined by the poles of the reflection coefficient $r_{\rm pp}$ of the p-polarized waves which are reflected into p-polarized waves. In Fig.~\ref{Fig:ReflectionC} we show the reflection coefficient for different frequencies $\omega$ and wave vectors $k_x$ choosing $k_y = 0$. To be more specific, we plot $\Im(r_{pp})$ in the evanescent wave regime ($|k_x| > \omega/c$) and $1 -|r_{\rm pp}|^2$ in the propagating wave regime ($|k_x| < \omega/c$). It can be nicely observed that there is a Zeeman splitting of the surface mode dispersion relation so that the surface waves in forward direction ($k_x > 0$) and backward direction ($k_x < 0$) fulfill different Zeeman shifted dispersion relations. This means that the heat carried by these waves at a given frequency is different in forward and backward direction which is the key behind the large heat flux rectification between the NPs coupled to such non-reciprocal surface waves in Refs.~\cite{OttSpin2020}. Here, this non-reciprocal behaviour can also explain the observed heat flux rectification in our SC. To this end, it is worth to note that there is a spin-momentum locking for the surface waves~\cite{Mechelen} so that the surface waves travelling to the left ($k_x < 0$) have a spin in positive $y$ direction and surface waves travelling to the right ($k_x > 0$) have a spin in negative y direction. As discussed for near-field thermal radiation in Ref.~\cite{OttSpin2020}, the coupling of the localized NP resonances and the surface modes is ideal when both have opposite spin. In Fig.~\ref{Fig:ReflectionC} we indicate the preferred coupling of the NP resonances with the surface waves by the encircled crossing of the corresponding NP frequencies with the surface mode dispersion relation. In the SC the NP resonances in both NPs couple the surface waves travelling in the same direction which are short range surface modes. In the AC the coupling becomes asymmetric and NP $1$ can efficiently couple to the long-range surface waves whereas NP $2$ couples to the short-range surface waves which is the reason for the observed stronger heating effect observed for NP $1$ in the AC.

%
%

\begin{figure}
	\centering
	\includegraphics[width = 0.5\textwidth]{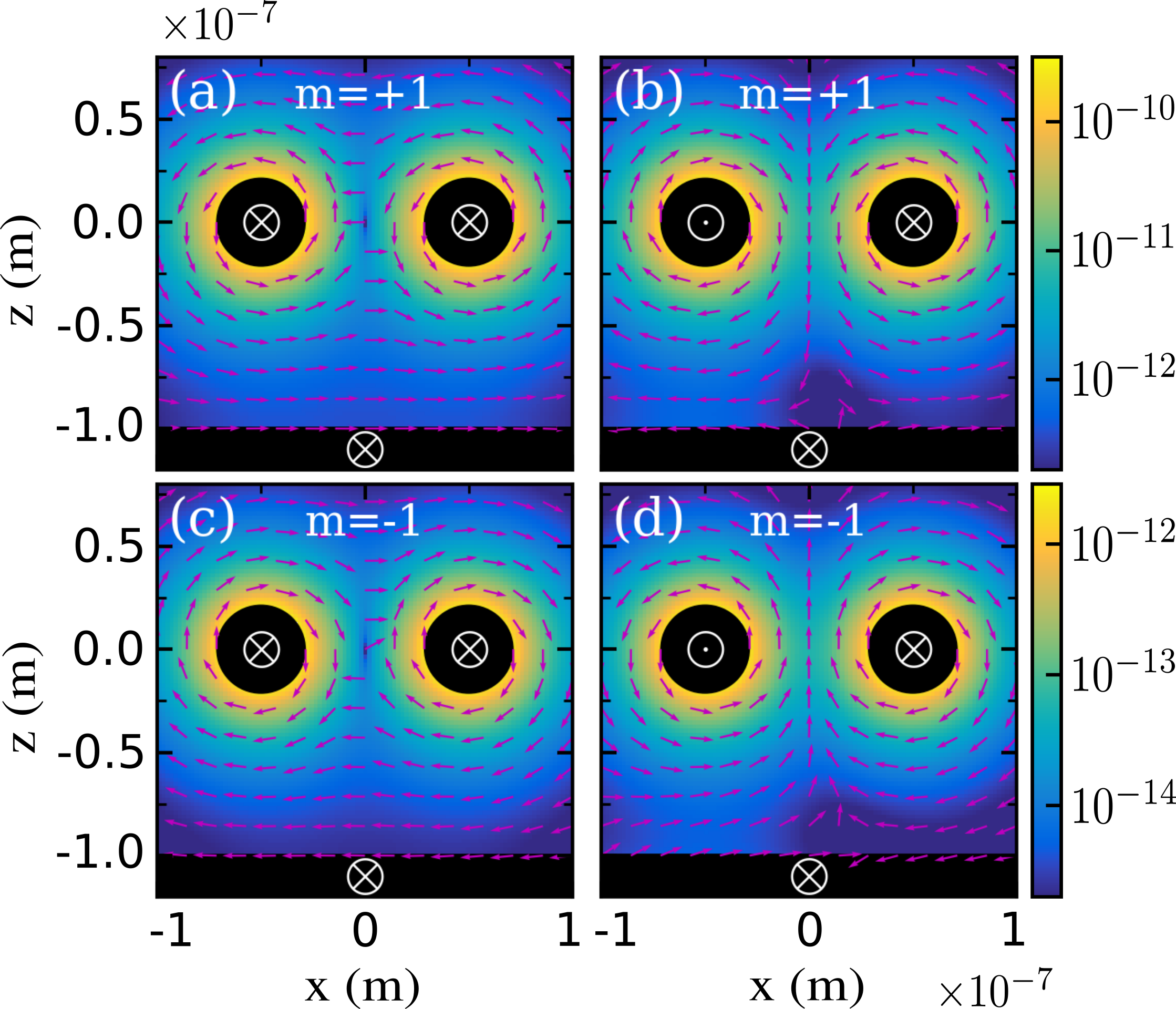}
	\caption{Spectral mean Poynting vector (Ws/m$^2$) from Eq.~(\ref{poyntingsurf}) for $T_1 = T_2 = 80\,{\rm K}$ and $T_b = 100\,{\rm K}$ evaluated at the NP resonances $\omega_{m = \pm 1}$ for the SC (left) and the AC (right).}
	\label{Fig:PV}
\end{figure}

To clearly demonstrate the spin-spin coupling mechanism in our system we calculate the mean Poyntingvector
in the vicinity of the two WSM NPs which are placed close to the WSM substrate. Because we are interested
in the part of the mean Poyntingvector which is responsible for the heat exchange we only plot the terms which
involve temperature differences ($\alpha = x,y,z$)~\cite{OttSpin2020}
\begin{equation}
\begin{split}
	\langle S_{\omega,\alpha}^{\rm tr} \rangle &= 4\hbar\omega^2\mu_0k_0^2 \sum_{\beta,\gamma = x,y,z}\epsilon_{\alpha\beta\gamma}\sum_{ijk=1}^{2}(n_j-n_b)\\
	                                           &\qquad  \times{\rm Re}\Big[ \mathds{G}_{0i}\blockt_{ij} \chimat_{j}(\mathds{G}^{\rm H}_{0k}\blockt_{kj})^\dagger\Big]_{\beta\gamma}. 
\end{split}
\label{poyntingsurf} 
\end{equation}
Here $\mathds{G}_{0i} = \mathds{G}(\mathbf{r},\mathbf{r}_i)$ is the electrical Green function and $\mathds{G}^{\rm H}_{0k} = \mathds{G}^{\rm H}(\mathbf{r},\mathbf{r}_k) = \nabla\times\mathds{G}(\mathbf{r},\mathbf{r}_k)/(\ri \omega \mu_0)$ is the magnetic Green function, $\mu_0$ is the permeability of vacuum, $k_0 = \omega/c$ is the wave number in vacuum,
\begin{equation}
  \chimat_{j} \approx \frac{\alphamat_j-\alphamat^\dagger_j}{2\ri} 
\end{equation}
is the susceptibility of the NPs $j = 1,2$ neglecting the radiation correction, and the T matrix expressing the interaction between both NPs with the substrate is defined by
\begin{equation}
  \boldsymbol{T}_{ij} = \delta_{ij}\mathds{1}-(1-\delta_{ij})k_0^2\alphamat\mathds{G}_{ij}. 
\end{equation}
The other terms of the mean Poynting vector which persist in global equilibrium~\cite{OttSpin2020} are neglected because they are irrelevant for the heat transfer in our system. 

In Fig.~\ref{Fig:PV}(b) and (d) it can be nicely seen that for the two frequencies $\omega_{m = \pm 1}$ there is a strong coupling effect for the NP and the surface modes of the substrate when the spin of the NP is opposed to the spin of the surface wave whereas when the spin in the NPs is aligned with the spin of the surface mode then there is only a weak coupling. Therefore one can expect a strong heat transfer from the surface to NP $1$ in the AC and a weak heat transfer to NP $2$ as also observed in the spectra for the received power in Fig.~\ref{Fig:P1P2}. Such a spin-spin coupling effect was shown in Ref.~\cite{OttSpin2020} to result in strong heat flux rectification between the two NPs and can be employed for thermal routing as shown in Refs.~\cite{GuoEtAl2020}. Here we can conclude that we have an asymmetric heat flux due to this spin-spin coupling of localized NP resonances and the surface modes of the substrate in the AC but not in the SC.

%
%

\begin{figure}
	\centering
	\includegraphics[width = 0.5\textwidth]{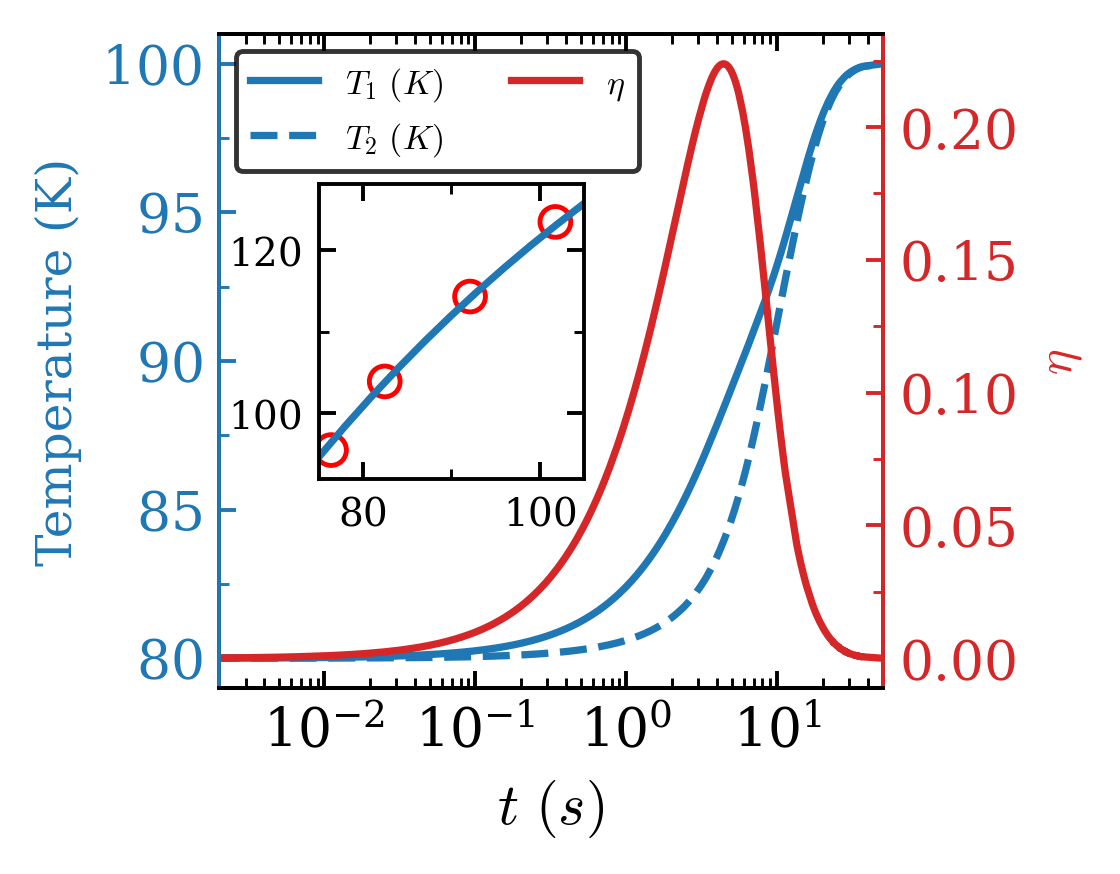}
	\caption{Dynamical evolution of the temperatures $T_1$ and $T_2$ of NPs $1$ and $2$ and the relative non-reciprocal heating efficiency $\eta = (T_1 - T_2)/\Delta T$ for the AC. The initial temperatures are $T_1 = T_2 = 80\,{\rm K}$ and the substrate temperature is fixed to $T_b = 100\,{\rm K}$. Inset: Experimental results (circles) for molar heat capacity $C_m$ (J/K mol) of Co$_3$Sn$_2$S$_2$ in the temperature range of $80\,{\rm K}$ to $100\,{\rm K}$ from Ref.~\cite{heatcapacity} together with the fit (solid blue line) used in our simulations.}
	\label{Fig:temp}
\end{figure}

Finally, we want to demonstrate the efficiency of the dynamical thermal routing  by employing Newton's law of cooling~\cite{Tschikin}
\begin{equation}
  \rho C_p V \frac{dT_i}{dt} = \langle P_{i}(t,T_1, T_2,T_b)\rangle, 
\label{dgltemp}
\end{equation} 
for both NPs with $i = 1,2$. The specific heat capacity $C_p$ of Co$_3$Sn$_2$S$_2$ is strongly temperature dependent and we use a fit to experimental results of the molar heat capacity $C_m$ from Ref.~\cite{heatcapacity} as shown in the inset of Fig.~\ref{Fig:temp}. For the molar volume we take $V_m = 69\times10^{-6}\,{\rm m}^3/{\rm mol}$ so that $C_p \rho = C_m/V_m = 1.46\times10^6 \, {\rm J} {\rm K}^{-1} {\rm m}^{-3}$ at $T = 80\,{\rm K}$, for instance. We start with the initial temperatures $T_b = 100\,{\rm K}$ and $T_1 = T_2 = 80\,{\rm K}$ so that $\Delta T = 20\,{\rm K}$. We are clearly in the temperature range below $T_c = 175\,{\rm K}$ where Co$_3$Sn$_2$S$_2$ has magnetic order and the WSM is non-reciprocal. Instead of using an eigenvalue~\cite{SanderEtAl2021} or response method~\cite{Naeimi2021} we solve Eq.~(\ref{dgltemp}) numerically with a Runga-Kutta method. As can be seen in Fig.~\ref{Fig:temp} the heating dynamics is very slow due to the fact that the value of the transferred power is relatively small. Furthermore, we see that for long times the NPs equilibrate to the temperature of the substrate, as expected. Nonetheless in the transient regime we observe a relatively big temperature difference between NP $1$ and $2$ of about $4.5\,{\rm K}$ which is the dynamical routing effect. Hence, we have a relative non-reciprocal heating efficiency $\eta = (T_1 - T_2)/\Delta T$ of $22.5\%$ which is much larger than the Hall effect of thermal radiation which is on the order of $1-2\%$~\cite{OttReview2019,OttAnomalous2020} or the non-reciprocal heating due to surface waves which is around $15\%$~\cite{OttSpin2020}. This result for our simple NP configuration shows that the non-reciprocal heating can have an important impact at the nanoscale and that WSM are good candidates to realize the thermal routing effect in nanoscale setups without the need of applying external magnetic field.  

%
%

To summarize, we have theoretically demonstrated the dynamical thermal routing of near-field heat radiation with the WSM Co$_3$Sn$_2$S$_2$. We find that there is a strong heat flux rectification due to the spin-spin coupling of the localized NP resonances and the surface modes of the substrate in the AC which results in a non-reciprocal dynamical heating of the NPs. The observed temperatures of the NPs show a $22.5\%$ difference with respect to the initially applied temperature difference. We are convinced that this effect cannot only be optimized for different configurations and materials but that our work also paves the way to an experimental realization in a three body measurement where the two NPs are replaced by either membranes as in Ref.~\cite{3body} or by nanowires, for instance.

%
%

The authors gratefully acknowledge financial support from the Niedersächsische Ministerium für Kultur und Wissenschaft (`DyNano').

\end{document}